\documentclass{article}

\usepackage{arxiv}

\usepackage[utf8]{inputenc} 
\usepackage[T1]{fontenc}    
\usepackage{hyperref}       
\usepackage{url}            
\usepackage{booktabs}       
\usepackage{amsfonts}       
\usepackage{nicefrac}       
\usepackage{microtype}      
\usepackage{lipsum}
\usepackage{subfiles}
\usepackage{subcaption}
\usepackage{graphicx}   
\usepackage[shortlabels]{enumitem}

\title{Multimodal Systems: Taxonomy, Methods, and Challenges}

\author{
  Muhammad Z.~Baig \\
  Department of Computing\\
  Macquarie University\\
  North Ryde, NSW 2109 \\
  \texttt{muhammad.baig@hdr.mq.edu.au} \\
   \And
 Manolya Kavakli \\
  Department of Computing\\
  Macquarie University\\
  North Ryde, NSW 2109 \\
  \texttt{manolya.kavakli@mq.edu.au} \\
}

\begin{document}
\maketitle

\begin{abstract}
Naturally, humans use multiple modalities to convey information. The modalities are processed both sequentially and in parallel for communication in the human brain, this changes when humans interact with computers. Empowering computers with the capability to process input multimodally is a major domain of investigation in Human-Computer Interaction (HCI). The advancement in technology (powerful mobile devices, advanced sensors, new ways of output, etc.) has opened up new gateways for researchers to design systems that allow multimodal interaction. It is a matter of time when the multimodal inputs will overtake the traditional ways of interactions. The paper provides an introduction to the domain of multimodal systems, explain a brief history, describe advantages of multimodal systems over unimodal systems, and discuss various modalities. The input modeling, fusion, and data collection were discussed. Finally, the challenges in the multimodal systems research were listed. The analysis of the literature showed that multimodal interface systems improves the task completion rate and reduces the errors compared to unimodal systems. The commonly used inputs for multimodal interaction are speech and gestures. In the case of multimodal inputs, late integration of input modalities is preferred by researchers because it allows easy update of modalities and corresponding vocabularies.
\end{abstract}


\section{Introduction}

The interaction between human and the world is multi-modal \cite{quek2002multimodal}. Humans utilize multiple senses to get an understanding of the environment. All the available senses are employed, both in series and in parallel, To continuously explore the environment and to perceive new information about the environment. The senses used in exploring the outside world can be sight, touch, listen, taste, and smell. These senses when used in a collaborative manner give the humans some useful insights about the surrounding - for example, sight is used to see an object, touch to identify the material, hearing to identify the source of the sound and estimate the location. Multiple modalities support humans to interact with the environment and other human beings effectively.

On the opposite side to human sensing techniques to interact which are inherently multi-modal, the human-computer interaction (HCI) techniques are primarily uni-modal - e.g., writing text using a keyboard. Multimodal interaction research goal is to design and develop interfaces, and technologies that eliminate the limitations of HCI and unlock its full potential. The introduction of new and sophisticated recognition algorithms enables the use of speech, gestures, and other modalities in the development of multi-modal interaction systems \cite{cutugno2012multimodal}. While the chances are highly unlikely that a multimodal interface completely replaces the traditional desktop or Graphical User Interface (GUI), but technological advancements are increasing the importance of MMIS.

A multimodal interface system (MMIS) gives an interface that bears the functionalities of human-human interface. Multi-modal interaction is achieved by combining the contribution of several research areas, including signal processing, computer vision, artificial intelligence, and many others. The overreaching goal of developing MMIS is increase the usability of computing technology. To improve interface usability, three things need to be studied: the user, the system, and the interaction.

 \section{Overview of Multimodal Interaction}

An MMIS conglomerates multiple input modes to achieve the multimedia output. These input modes functioned in a coordinated manner. Speech, gestures, pen, touch, and gaze movement are the commonly used user input modes \cite{Oviatt2003}. Several studies have used human behavior and a common language to interact with computer \cite{fariman2016designing}. These MMIS utilize recognition algorithms to understand and describe the language and behavior of the user. According to Oviatt \cite{Oviatt2003}, \textit{" Multimodal interfaces process two or more combined user input modes (such as speech, pen, touch, manual gesture, gaze, and head and body movements) in a coordinated manner with multimedia system output. They are a new class of interfaces that aim to recognize naturally occurring forms of human language and behavior, which incorporate one or more recognition-based technologies (e.g., speech, pen, vision)".}

The term "multimodal" could be used in various contexts; In HCI, multimodal is defined using a more human-centered approach. Modality is the mode of communication used as input to activate the computer, and it is a measure of human senses and actions. The input modalities that resembles the human senses are cameras (sight), microphones (hearing), haptic (touching) \cite{hafez2007tactile}, olfactory (smell) \cite{liu2012schedule} and electronic tongue (taste) \cite{riul2010recent}. The biofeedback input devices such as skin conductance, heart rate, Electroencephalogram (EEG) and many other, used to measure the internal activity of humans, are also considered as input modalities \cite{turk2014multimodal}.

In HCI, the most common interfaces are perceptual interfaces, attentive interfaces, and enactive interfaces. A brief definition of these interfaces is given below: 

\begin{itemize}
\item Perceptual interfaces provide natural, rich, and efficient interaction with the computer using multimodal inputs, and it is highly interactive \cite{turk2004perceptual}. 
\item Attentive user interfaces are the ones that depend on user's attention and use the gathered information from modalities to approximate the suitable time for interacting with the user \cite{chen2004using}. Many applications of attentive user interfaces involve computer vision as the main component to perform several functions such as eye tracking, facial emotion, and gestures \cite{bostandjiev2012tasteweights}. 
\item Enactive interfaces are those interfaces that allow the expression and communication of enactive knowledge to actively utilize the use of hands and body for understanding task \cite{bennett2007towards}.
\end{itemize}

The most widely used input mode in commercial application is speech. Nowadays, speech-controlled assistant is a must include functionality in the smartphones. Gestures for interaction have shown some promising practical application as well \cite{Erol2007}. However, people preferred multiple input modes over uni modal input for interaction because it improves the handling, reliability, and task-completion rate of the system \cite{Oviatt1997}. Table \ref{tab:table} describes some attributes that differentiate a traditional unimodal interface from a multimodal interface \cite{Dumas2009}.
\begin{table}[h]
\centering
\caption{Traditional systems vs MMIS \cite{Dumas2009}}
\renewcommand{\arraystretch}{1.3}
\begin{tabular}{|p{1.7in}|p{2.2in}|} \hline 
\textbf{Traditional System}s & \textbf{Multimodal Interface System} \\ \hline \hline
unimodal input & multimodal input \\  
atomistic, deterministic & continuous, probabilistic \\  
sequential processing & parallel processing \\ 
centralized architectures & distributed and time-sensitive architectures \\ \hline
\end{tabular}
\label{tab:table}
\end{table}

 \section{History of Multimodal interface system}
\begin{figure}[ht]
\centering
\includegraphics[width=0.5\linewidth]{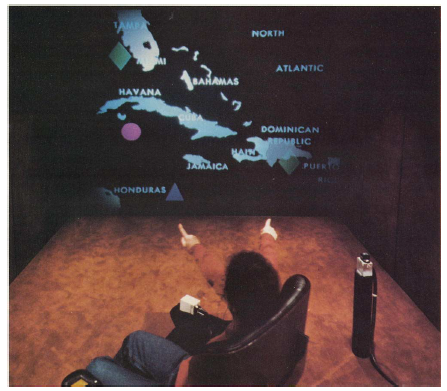}
\caption{Bolt's "Put that there" MMIS \cite{bolt1980put}}
\label{fig2}
\end{figure}

An MMIS typically comprises of a recognition system. Recognition system function is to translate the human activities into identifiable computer signals. In the next step, the system decodes the input and combine it with other modalities to attain the required output. There exist many MMIS that uses speech, gestures, and pen input  \cite{Kavakli2012}. Bolt's "Put That There" MMIS is one of the earliest implementation in which speech and pointing gestures were aggregated to move an object \cite{bolt1980put}. Fig. \ref{fig2} shows an interface of the "Put That There" experiment. This experiment is considered a groundbreaking demonstration of multimodal interfaces.
After "Put That There" experiment, the multimodal inputs, especially speech and gesture, are used in various applications. The early multimodal system was mainly used to perform the spatial task in a map-based environment. Neal et al. developed a multimodal system CUBRICON for tactical mission planning that uses speech, typed text, and gesture as input and displayed the output using a combination of language, maps, and graphics \cite{neal1989natural}. Koons et al. also implemented an MMIS for a map-based application that uses speech and gesture for interaction \cite{koons1993integrating}. In 1997, Cohen et al. developed Quickset, an MMIS that uses pen/voice, as a training simulator for US Marine Corps \cite{cohen1997quickset}. Speech combined with gesture input has also utilized to draw sketches  \cite{Newman2008}. Kinect and Leap motion sensors are commonly used to recognize gesture inputs. The Speech functions as an add-on for the user such as changing texture or orientation of the model \cite{Rodger2004}.

The multimodal interaction considered as the expansion of traditional desktop experience, but a decent amount of research also focused on substitute or "post-WIMP" computing environment which brings new input modalities such as touch and haptics interfaces \cite{turk2014multimodal}. Post WIMP interfaces are those interfaces that are far away from traditional graphical user interfaces and rely on speech and gesture \cite{van1997post}. These Post WIMP interfaces include a more robust "butler-like" interface that gives life to a new generation of perceptual interfaces \cite{turk2004perceptual}. These perceptual user interfaces use a more natural way by incorporating the human capabilities such as communication, cognition, motor skills, and perception into the interface system.

 \section{Differences Between Unimodal and Multimodal Systems}
Usually, MMIS intend to deliver a natural and efficient interaction between human and computer, as there are some advantages of using a multimodal system over a unimodal system. The literature on the assessment of MMIS state that users prefer multimodal systems over unimodal systems \cite{oviatt2006human}. Multimodal systems also provide flexibility and reliability \cite{xiao2002multimodal,oviatt2005individual,bohus2010facilitating} and increase task efficiency. Information processing improves when the information is presented using multiple modalities \cite{Dumas2009,van2005visual}. Other possible advantages of multimodal interaction systems explained by \cite{Oviatt2000} include:

\begin{itemize}
\item MMIS allows nonrigid use of modalities, including sequential and parallel use.
\item MMIS improve system efficiency, provide greater precision of spatial information, and bring robustness to the interface.
\item MMIS gives user alternatives in interaction, enhance the error avoidance and correction mechanism.
\item MMIS can be made adaptable for a continuously changing environment and accommodate individual differences.
\end{itemize}

MMIS was considered more effective than the traditional (unimodal) interfaces, but evaluations studies showed that MMIS improve the task completion rate by only 10\%  \cite{Dumas2009}. On the other hand, multimodal interfaces reduce errors by 36\% compared to unimodal interfaces. 
Despite so many advantages, it is hard to extrapolate the conclusions as for every sequence of task, user, environment, and the required interface is significantly different. Sometimes the use of multiple inputs may degrade the performance \cite{oviatt1999ten}.

 \section{Modality}
In different fields, the terms relevant to MMIS such as multimodal, modality, devices, and multimedia  means different things. \cite{turk2014multimodal}. In HCI terms, Modality is the form in which information is displayed or transferred, such as speech, text, visual, gestures, etc. Each input is transferred to a computer by a specific medium. For example, the text is entered through a keyboard and visual information through a camera. Different modalities have different definitions based on their properties and representation. Modalities are also information-dependent, which means that some particular type of modalities could be suited for one particular type of information but not for other types \cite{Cao2011}.

\subsection{Input and output modalities}

An MMIS can respond to multiple input modalities such as speech, gaze, gestures, etc. in an organized way to achieve a particular output. MMIS become a new focus of interest for the future computing generation since they have shifted the paradigm away from the standard keyboard mouse input.
The earliest examples of MMIS were probably the ones that were least different from traditional Graphical User Interface (GUI) systems as they only reduced the use of keyboard and mouse as input modes. Since speech and gesture recognition technology has very much matured, the typical GUI systems utilize speech and gesture inputs along with standard keyboard and mouse interfaces to address user intention \cite{Liu2014}.

Blattner and Glintert \cite{blattner1996multimodal} listed some input and output modalities and examples. The list was further updated by Turk et al. \cite{turk2014multimodal} given in Table \ref{tab51}. We have included some bio-sensor modalities to the list as well because these modalities can be an integral part of future Multimodal systems.
\begin{table}[t]
    \centering
    \begin{tabular}{|l|l|}\hline
        \textbf{Modalities} &    \textbf{Examples} \\ \hline \hline
Visual&    Face localization \\
&Gaze location\\
&Facial expressions\\
&Lipreading\\
&Face detection \\
& Gestures (head/face, hands, body) \\
& Sign language\\ \hline
Auditory &    Speech input\\
&Non-speech audio\\ \hline
Touch    &Pressure\\
&Location and selection\\
&Gesture\\ \hline
Bio-sensors & Brain-computer interfaces\\
& Emotion recognition\\
& Cognitive load estimation\\ \hline
Others &    Motion capture \\ \hline
    \end{tabular}
    \caption{Sensor modalities in Multimodal interaction system \cite{blattner1996multimodal,turk2014multimodal}.}
    \label{tab51}
\end{table}

A human perceives the environment by utilizing the five available senses such as smell, sight, touch, taste, and hearing. The pathway through which the information from the senses is transmitted or received is called communication channel \cite{Dawson2007}. In multimodal interaction, a channel is an interaction technique through which humans transmit or receive information based on user ability and device capability \cite{pantic2003toward}. For example, a keyboard is for text input, mouse for pointing or selecting input or camera for visual input. 

Several key factors and characteristics constitute multimodal system architecture and development. These dimensions or characteristics are \cite{Oviatt2003}:
\begin{itemize}
\item Input modalities (size and type)
\item Communication channel (devices and size)
\item Processing modes (series or parallel or both)
\item Vocabulary (size and type)
\item Sensors and channel fusion
\item Application
\end{itemize}

In a multimodal interaction system, figuring out the correct characteristics for the MMIS architecture is not simple, the designer needs to take the decisions without the use of rational processes or by testing but to find out the best multimodal input for an interface is still an open research question \cite{kim2016multimodal}. Most of the work in multimodal interaction focuses on recognizing modalities such as gesture, speech, and facial expression recognition. A few studies focus on the output modality, medium of sensory output between human and a machine, which is also a key element of HCI \cite{obrenovic2004modeling}. With the availability of powerful mobile devices (smartphone) and embedded sensors such as the microphone and 3D vision sensor (Kinect, leap-motion, 3D scanners, and printers) has opened abundance of possibilities for MMIS and the HCI world.

\subsection{Common Multimodal Inputs}
With the advances in software and hardware technologies, the MMIS design has become an interesting research domain in recent times. Nowadays, people preferred to use natural and human-like ways to interact with computers \cite{begic2000quantitative}. It also enables the researchers to integrate inputs in series or in parallel to create new set of modalities such as speech and pen input \cite{shinoda2011semi,ruiz2010cognitive,dusan2003multimodal}, speech and gestures \cite{miki2014improvement,ismail2015multimodal}, and speech and lip movement \cite{wang2012sentence,stork2013speechreading}. Nowadays, speech and gestures are the most commonly used modalities in multimodal interaction.
\subsubsection{Speech}
The most common input modality for a multimodal interface is speech. Speech is considered a medium for communication, and it consists of words which are further categorized into grammar, syntax, semantics, discourse, pragmatics, and prosody. These parts help us understand the language.

\begin{itemize}
\item \textbf{Grammar} is used to make rules and laws, and the right methods to apply these for speaking and writing. 
\item \textbf{Syntax} links names and actions together in a defined order. 
\item \textbf{Semantics} involves the study of the meanings of individual words and syntactic contexts.
\item \textbf{discourse} is written and spoken communication that constitutes a sequence of relations to the subject, object and announcements. 
\item \textbf{Pragmatics} is defined as how language is used and investigates the semantic and syntactic uses of language. 
\item \textbf{prosody} is the emotion state-of-utterance \cite{Rabiner1993}.
\end{itemize}
\paragraph{Speech Recognition}
Speech recognition is used to translate speech into text.  Various recognition algorithms are available which can be categorized in three main types: Hidden Markov models (HMM) \cite{baker2009}, Dynamic time warping (DTW) \cite{muda2010}, and Neural Networks \cite{maas2012}.  Microsoft (MS) compiled an API in 1994 named "Speech Recognition and Synthesis API" which has the capability to translate speech into text and vice versa in runtime. The API can be used to convert various languages including English, Spanish, Chinese, etc. \cite{Zuberec2001}. MS  API claimed to have an average speech recognition rate of more than 75\% \cite{burger2006competitive}. There are some other speech recognition systems such as Carnegie Mellon University (CMU) Sphinx-4 based on HMM, and Google API based on deep neural networks that are the possible alternatives to MS speech recognition API which provide better performance \cite{kepuska2017comparing}. 
\begin{figure*}[t]
\begin{subfigure}{0.4\textwidth}
\centering
\includegraphics[height=4cm]{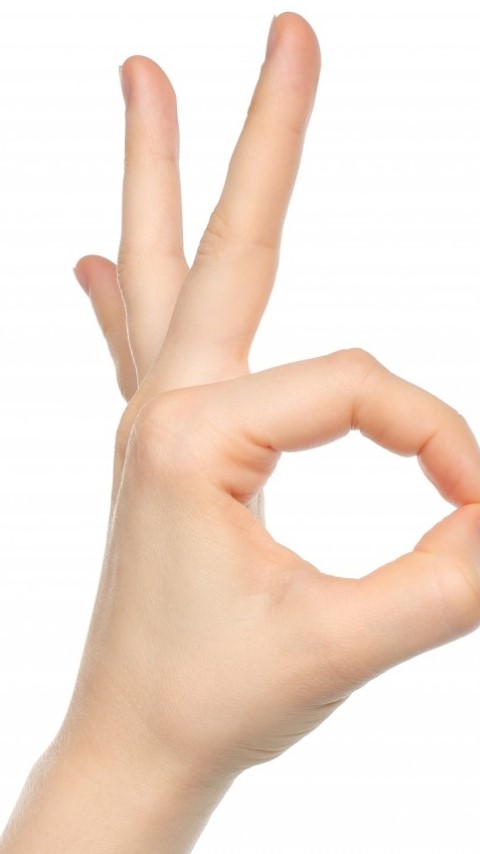}
\caption{Symbolic gesture "OK"}
\label{fig252a}
\end{subfigure}
\begin{subfigure}{0.4\textwidth}
\centering
\includegraphics[height=4cm]{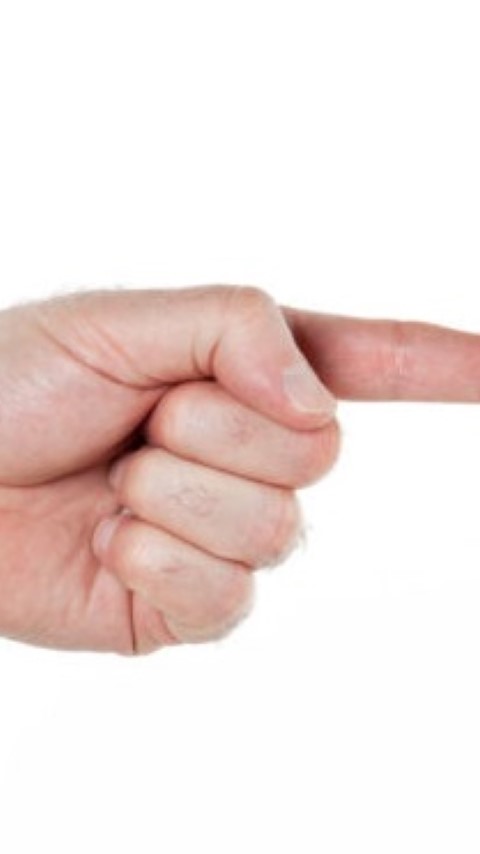}
\caption{Deictic gesture "Pointing"}
\label{fig252b}
\end{subfigure}
\begin{subfigure}{0.5\textwidth}
\centering
\includegraphics[width=0.8\linewidth]{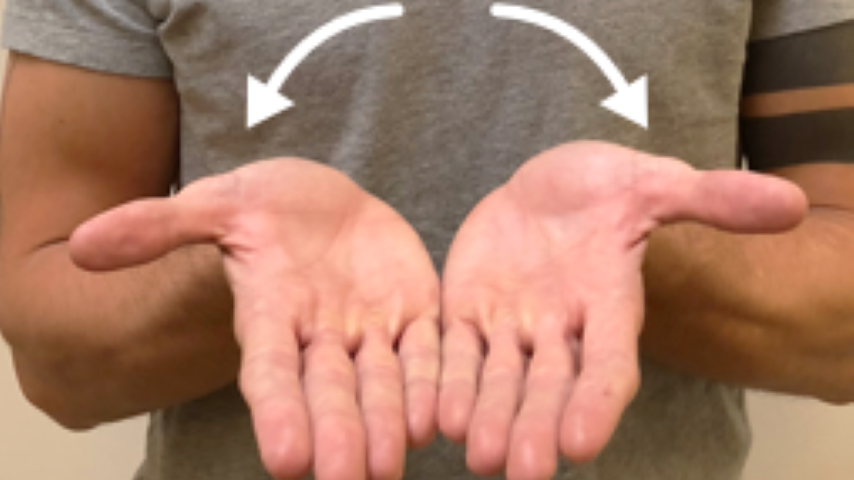}
\caption{Iconic gesture "Book"}
\label{fig252c}
\end{subfigure}
\begin{subfigure}{0.5\textwidth}
\centering
\includegraphics[width=0.8\linewidth]{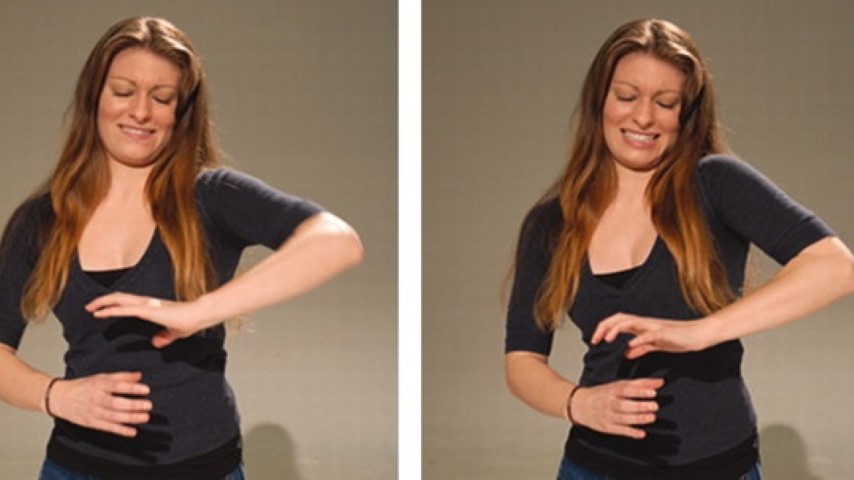}
\caption{Pantomimic gesture "opening a jar"}
\label{fig252d}
\end{subfigure}
\caption{Gestures examples}
\label{figtheta}
\end{figure*}
\subsubsection{Gesture}
The movement of the body part to convey a message is known as a gesture. However, in the literature, gestures are hand and head movements. Many applications exist in the literature that utilized gesture for interaction \cite{Vafaei2013}. The term gesture first appeared in 1979 in a book named "GESTURES, their Origins and Distribution" by Morris et al. \cite{morris1979gestures}. In this book, an analysis of emblem gestures, the action that replaces speech, was given. Rime and Schiaratura explained a gesture taxonomy for communication with a computer \cite{Rime1991}. These gestures are symbolic, deictic, iconic and pantomimic gestures.
\begin{itemize}
\item \textbf{Symbolic gestures} are those gestures that have a single meaning. Fig. \ref{fig252a} shows the symbolic gesture of "OK".
\item \textbf{Deictic gestures} are the pointing gestures as shown in Fig. \ref{fig252b}. These gesture are used to tell the other person about specific object or event in the surrounding environment.
\item \textbf{Iconic gestures} give the information of an object. The information includes size, shape, or orientation of an object. For example, a person doing gesture to describe a rectangular object as shown in Fig. \ref{fig252c}.
\item \textbf{Pantomimic gestures} are used  to show the movement of some invisible tool. For example, turn a knob as shown in Fig. \ref{fig252d}.
\end{itemize}

McNeill in 1992 added two more gesture to these taxonomies that relate to the process of communication \cite{Billinghurst2011}. These gestures are beat and cohesive gestures:
\begin{itemize}
    \item \textbf{Beat gestures} are used to indicate a pace, represent the up and down movement with the rhythm of speech.
    \item \textbf{Cohesive gestures} are used to combine temporally separated, but thematically relation portions of discourse, and these are the variation of iconic, pantomimic or deictic gestures.
\end{itemize}
 Cadoz in 1994 associated three types of functions to the group of gestures: semiotic gestures, ergotic gestures, and epistemic gestures \cite{Billinghurst2011}.  
 \begin{itemize}
     \item \textbf{Semiotic gestures} "are used to communicate meaningful information."
     \item \textbf{Ergotic gestures} "are used to manipulate the physical world and create artifacts."
     \item \textbf{Epistemic gestures} "are used to learn the environment through exploration."
 \end{itemize}
 Later on, in 2005, Karam et al. \cite{karam2005taxonomy} provide a comprehensive taxonomy of gesture in HCI. They categorize the gestures in deictic, semaphores, gesticulation, manipulation, and sign language gestures. In 2009, Wobbrock et al. explained a taxonomy for surface gestures based on the evaluation of how many users performed a particular gesture for a given task \cite{wobbrock2009user}. Ruiz et al. in 2011 proposed a taxonomy for 3D motion gesture and applied it on smartphones \cite{ruiz2011user}. They used the physical characteristic such as kinematic impulse, dimensionality, and complexity to create a user-defined gestures dataset.
\paragraph{Gesture recognition}
Through the years, gesture recognition has been through tremendous advancement. Three types of algorithms existed in the literature for gesture recognition The first type is \textit{\textbf{glove-based methods}}. The glove-based gesture algorithms are further have two techniques: active and passive data gloves. To estimate joint twisting and acceleration, embedded sensors are used that are mounted on the gloves. Vision based approach is used in the passive data gloves to measure the location and angle of the hand and fingers. This technique uses color pointers for finger detection \cite{Premaratne2014}. 

\textit{\textbf{Haptics}} is the second common technique for gesture recognition. In air tactile experience has been utilized for recognizing gestures. The advantage of this technique is that the subject is not asked to wear any sensor device. The most common example of haptics technology for gesture recognition is AIREAL. In air tactile experiences are extracted using a directed air vortex generation in AIREAL. AIREAL provides decent tactile feedback with a 75-degree field of view, and an 8.5 cm resolution at 1 meter \cite{Sodhi2013}.

The third technology for recognizing gesture is \textit{\textbf{sensor-based}} motion capture. In sensor-based motion capturing, sensors raw data are processed to measure the location and type of gesture. The most commonly used sensors  are Microsoft Kinect and Leap motion \cite{molchanov2016online}. 


\section{Multimodal interface system development}
Development of the multimodal system is challenging because traditional computing techniques usually do not applied into a multimodal environment effectively \cite{turk2014multimodal}. In addition to that, each factor or characteristic of the multimodal system may result in different design strategies. Oviatt in 1999 \cite{oviatt1999ten} proposed a set of myths about the multimodal systems "Ten Myths of Multimodal Interaction" have proven to be useful in the MMIS field. These ten myths, as mentioned in \cite{oviatt1999ten} are given below:

\begin{enumerate}[start=1,label={\bfseries  \arabic*:},wide = 0pt, leftmargin = 4em]
\item "If you build a multimodal system, users will interact multimodally."
\item "Speech and pointing is the dominant multimodal integration pattern."
\item "Multimodal input involves simultaneous signals."
\item "Speech is the primary input mode in any multimodal system that includes it."
\item "Multimodal language does not differ linguistically from a unimodal language."
\item "Multimodal integration involves redundancy of content between modes."
\item "Individual error-prone recognition technologies combine multimodally to produce even greater unreliability."
\item "All users' multimodal commands are integrated uniformly." 
\item "Different input modes can transmit comparable content."
\item "Enhanced efficiency is the main advantage of multimodal systems."
\end{enumerate}

In 2004, Reeves et. al. \cite{reeves2004guidelines} give some guidelines for MMIS design:

\begin{itemize}
\item \textbf{User Specifications:} The MMIS should be designed by keeping in mind a broad variety of users and contexts. Privacy and security issues should be addressed. 
\item \textbf{Input and Output Specifications:} The design should maximize the human cognitive and physical capabilities by considering the information processing abilities and limitations. The system modalities should be harmonious with the user preference, context, and system functionality.
\item \textbf{Adaptivity:} The multimodal system should adapt to the needs and capabilities of users.
\item \textbf{Consistency:} The system should be consistent in presentation and prompt.
\item \textbf{Feedback:} The users should know the recent connectivity and available modalities.
\item \textbf{Error Prevention/Handling:} The system should provide error detection, prevention, and handling functionalities.
\end{itemize}


\section{Modelling, Fusion, and Data Collection}
The basic principles and methods needed to develop a Graphical User Interface (GUI)-based interaction may not be used in MMIS development, which makes the multimodal interface design special \cite{reeves2004guidelines}. As mentioned earlier, special attention should be given to input/output modalities, adaptability, consistency, and error handling issues. The human-related factors such as personality, background, current emotional state need to be considered when designing a multimodal interface \cite{jaimes2006human,sebe2009multimodal}. These issues and design decision further prescribe the fundamental algorithms and methods used in the development of interfaces.
\subsection{System Integration Architectures}
In the multimodal community, Open Agent Architecture \cite{martin1999open} and Adaptive Agent Architecture \cite{cohen2004semisupervised} are the commonly used multi-agent architectures. Multi-agent architectures use a distributed approach to implement the complex modules of multimodal processing. The modules and components in a multi-agent architecture can be developed in various programming languages, different machines, and different operating systems. Inputs in a multi-agent architecture can be in parallel or asynchronous and then be passed to the recognition system. The results from the recognition modules interpreted and delivered to the user through multimedia feedback \cite{turunen2003jaspis}.

\subsubsection{Open Agent Architecture (OAA)}
OAA combines the functionalities of those agents who cannot work collectively, thereby help wider reusing of agent's expertise \cite{martin1999open}. The key attributes of this architecture are:
\begin{itemize}
\item \textbf{Open:} The OAA works with agents written in various languages and on variety of platforms. 
\item \textbf{Extensible:} Agents can be included or excluded from the system at run-time. 
\item \textbf{Mobile:} Low-end portable computers can be used to execute OAA-based applications.
\item \textbf{Collaborative:} The OAA consider the user as another agent which simplifies creating collaborative systems with multiple users and agents.
\item \textbf{Multiple Modalities:} In addition to the traditional GUI, OOA-based interface supports multiple modalities.
\item \textbf{Multimodal Interaction:} In OAA- based architecture, users can use multiple modalities to interact with the system.
\end{itemize}
\subsubsection{Adaptive Agent Architecture (AAA)}
AAA is a robust brokered (or middle agents) framework that "uses teamwork to recover a multi-agent system broker failures and to control a minimum limit of system's functional brokers even when some of the brokers become unreachable \cite{kumar2000adaptive}." To achieve the functionality AAA has two mission statements as mentioned in \cite{kumar2000adaptive}:

\begin{itemize}
\item \textbf{{Mission Statement 1:}} "Whenever an agent registers with the broker team, the brokers have a team intention of connecting with that agent, if it ever disconnects, as long as it remains registered with the team." 
\item \textbf{{Mission Statement 2:} } "The AAA broker team has a team maintenance goal of having at least N brokers in the team at all times where N is specified during the team formation."
\end{itemize}

\subsection{User Modelling and Human Information Processing}
Modeling human information processing in HCI and related fields is a challenging job, and several studies are focusing on this challenge received significant attention \cite{sebe2009multimodal,yoshikawa2002modeling,guo2018user}. In this section, we have mentioned some commonly used models. The most famous model in the HCI literature is the Model Human Processor \cite{card2018psychology}. It has three parts: the perceptual, cognitive, and motor system. The perceptual system is responsible for handling the sensory information from the outside world, i.e., the input-output components. The motor system controls all the actions, also known as a processing component. The cognitive system, on the other hand, provides the functionality to connect the other two systems \cite{jastrzembski2007model}. According to the model human processor, the input-output channels (movement, hearing, vision), memory (both long and short term) and the processing (problem-solving, reasoning) need to be considered when designing an MMIS \cite{dix2004human}.

GOMS (Goals, Operators, Methods, and Selection rules) is another famous architecture proposed by Card et al. \cite{card2018psychology}. There is now a family of many variations of GOMS. The model is suitable for modeling the optimal behavior of the users. Despite so many variations of GOMS, the fundamental concept is not different. \textbf{Goals} are the result that a subject wants to achieve. \textbf{Operator} is the action executed for goal fulfilling. The sequence of operators is called a \textbf{method}, and if there exists many different method, then a \textbf{selection rule} is used to shortlist one method \cite{biswas2010brief}. All GOMS variations provide useful insight, but on the cost of some drawbacks. For example, these algorithms do not incorporate the user fatigue into the model, and the user performance degrades over time because the user performs the same task repetitively. GOMS techniques are flexible to necessary cognitive actions and at times applicable to expert users.

There are some other models including the cognitive architecture models \cite{byrne2007cognitive} (ACT-R/ PM \cite{byrne2001act}, LICAI/CoLiDeS \cite{hijazi2004cognitive}), the psychological models such as human action cycle \cite{norman1988design,cahn1999psychological}, grammar-based models \cite{chase2002model} and application specific models \cite{motomura2000generative}.

\subsection{Adaptability}
Adaptability is a vital part of the current and future HCI systems because of the large, diverse types of users. The HCI system should accommodate for the difference in expertise, culture, language, and goals by introducing adaptive and customizable interfaces. The system needs to learn through user behavior and knowledge to predict the user's future behavior and response \cite{browne2016adaptive}. With the inclusion of adaptiveness in the system, it can improve the overall user performance and make an interesting interaction \cite{lavie2010benefits}.

The feedback element in adaptive HCI promises to increase the user ability to handle complicated tasks easily with fine accuracy and allows the user to learn new techniques quickly. The adaptive HCI interfaces provide an interactive knowledge or agent-based dialog to handle errors, interruptions, understand the current circumstances \cite{oppermann2017adaptive}. The far fetch aim of intelligent user interaction is to increase the effectiveness, naturalness, and productivity \cite{ross2000intelligent}. The challenging part in developing an adaptive user interface is to find the variables on which the system can adapt. The system needs to analyze input, user behavior, and actions and find those variables that maximize the efficiency, effectiveness of the interface \cite{pantic2005affective}.

Adaptive systems need continuous feedback from the user to learn, and for that reasons, these systems are also called learning systems. As these systems involve in a continues learning cycle, the developer should not consider the adaptive systems a solution for all the problems and focus on whether the user needs an adaptive system. Some researchers state that adaptive interfaces violate the standard usability principle, but sometimes a static user interface, don't depend on user state, shows superior performance over an adaptive interface \cite{hook2000steps,chin2001empirical}. Despite these studies, adaptive systems can bring undeniable benefits to the interaction system. Nowadays, many researchers work on developing intelligent systems form many different applications including learning and tutoring systems \cite{mahdi2016intelligent,alawar2017css,al2018intelligent}, medical applications \cite{sonntag2017overview}, smartphones \cite{lee2011adaptive,zhang2012nihao,lopez2017alexa} and arts \cite{el2007interaction}.

Although there are many advances in the adaptive interface systems, still there are various research problems that need answering such as usability of input modes, finding variables to adapt, and estimating user behavior.

\subsection{Multimodal Integration}
In daily communication, all our input modes are coordinated perfectly, e.g., speech, lip movement, hand gesture, and facial expressions. It is not the case with computers; the inputs in MMIS are not designed to coordinate at all with each other. Unlike humans, computers with an MMIS produce a challenge for integrating complementary modalities to generate a highly cooperative combination. Many fusion techniques and architectures were developed to integrate multi-modal inputs for joint processing \cite{Paleari2006}.

In older multimodal systems, the data collected through various modalities are processed separately and fuse in the end. Modalities with different characteristics such as speech and gesture cannot be connected straight away. Therefore, to integrate inputs with different characteristics, three different levels of integration, also known as fusion levels, are proposed. These levels are signal level, features level, and semantic level \cite{corradini2005multimodal,lalanne2009fusion}. Table \ref{tab81} summarizes these three fusion levels.
\begin{table}[h]
\centering
\caption{Summary of three fusion levels \cite{Paleari2006}}
\renewcommand{\arraystretch}{1.3}
\begin{tabular}{p{1in}||p{1.35in}|p{1.35in}|p{1.45in}} \hline 
\textbf{Fusion level} &\textbf{Signal-level fusion} & \textbf{Feature-level fusion} & \textbf{Semantic-level fusion} \\ \hline\hline
\textbf{Input type} &    Raw    signal of same type &    Closely synchronized & Loosely coupled \\ \hline 
\textbf{Level of information} &    Highest level of information detail &    Moderate level of information detail & Mutual disambiguation by combining data from modes \\  \hline
\textbf{Noise/failure sensitivity} &    Highly susceptible to noise or failures    & Less sensitive to noise or failures &    Highly resistant to noise or failures\\ \hline
\textbf{Utilization} &    Rarely used for fusing multiple modalities & Used for fusing specific modes &    Commonly used fusing technique \\ \hline
\end{tabular}
\label{tab81}
\end{table}
\subsubsection{Signal level fusion}
In signal level data fusion, the raw sensor data from different modalities are combined. This level is more suited for integrating highly synchronous inputs, such as speech and lip movements, otherwise the performance of the system got effected. The signals are combined in a vector form at a very early stage. The dimension of the vector is reduced by applying dimensionality reduction techniques (e.g. LDA, PCA) and then the vector is forward to the recognition engine for the classification \cite{Paleari2006}.

\subsubsection{Feature level fusion}

The problem with the signal level fusion is that it fails to model the fluctuations and the asynchrony between the input modalities. One solution is to extract features from the signals and use the feature to combine the modalities. Features based on transformations, probabilistic models, information measures, etc. can be used for fusion in this level \cite{sebe2009multimodal}. 
\subsubsection{Semantic fusion level}
The semantic level fusion of input modalities uses common meaning representation extracted from different modalities into a combined interpretation. Instead of working on raw signals or feature, the semantic fusion works on semantic information extracted from individual modalities and combine the interpretation from each mode. This fusion level can control loosely-coupled input modalities such as speech and gestures \cite{Liu2014}. Semantic level fusion has advantages over other fusion techniques such as the multiple inputs do not have to co-occur because each modality can be recognized independently \cite{turk2014multimodal}. 

Independent classifiers have been used in semantic fusion level, and the final decision is made by integrating unimodal classifiers output. The semantic level fusion provides the advantage of reducing computational complexity by training classifiers separately ($O(2N)$) instead of combined training ($O(N^2)$)  decreases the computational complexity \cite{turk2005multimodal}. The semantic fusion approach has been used widely for the coupling asynchronous input modes such as speech and pen or speech and gestures. It is easy to scale up the input modes and vocabulary size in semantic level fusion because of the late integration of modalities.

\subsection{Frameworks for Input Information  Fusion}
To fuse multimodal inputs, we need a framework. It is an important requirement in MMIS design. The framework requires a time stamp of individual inputs. In case of speech and gesture inputs, these can be used in a series or parallel. A timestamp is a vital flag to check the starting and ending of a multimodal input \cite{Oviatt2000}.
\begin{figure}[h]
    \centering
    \includegraphics[width=1\linewidth]{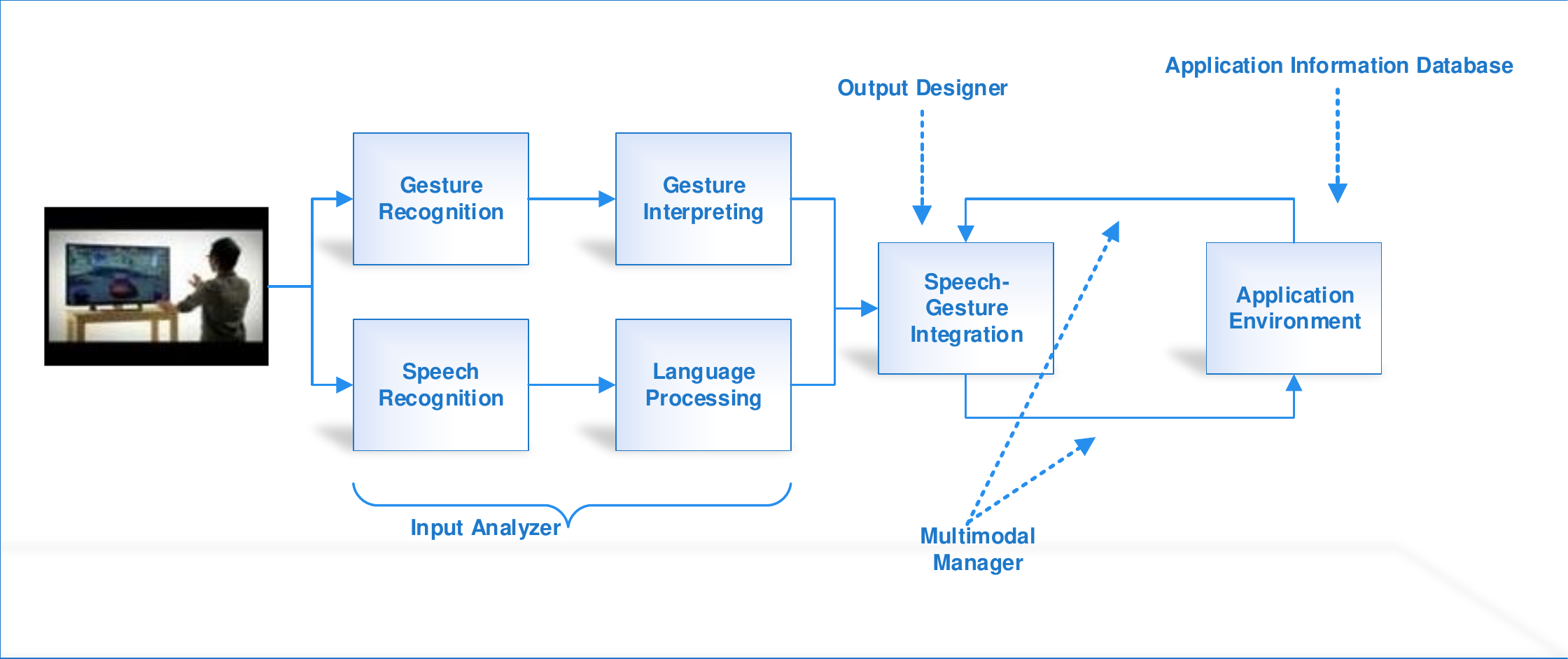}
    \caption{Framework for semantic level speech and gestures integration}
    \label{fig71}
\end{figure}
Fig. \ref{fig71} shows a common framework for speech and hand gesture input at the semantic level. The framework has four main parts: Input Analyzer, Multimodal Manager, Output Designer, and Application Information Database \cite{liu2010survey}. The input analyzer processes speech and gesture in parallel, and the results are represented in semantics that is given to the output designer block. The task of the multimodal manager is to exchange information between the output designer and application information database for real-time control. The problem with these kinds of architecture is the use of a similar programming language which can be difficult to use in some cases. To overcome this problem,  a multi-agent architecture is proposed \cite{Paleari2006} which uses multiple agents for pooling the pre-semantic information from different sensors and integrate them into a multimodal data structure.

\subsection{Data Collection and Testing}
In an MMIS, one of the most challenging tasks is to obtain the ground truth by collecting and labeling data as these are error-prone and take a lot of effort and time. Usually, in HCI, we rely on self-reported data from the user. For example, in emotion recognition, the data used is mostly based on simulated data by actors. Actors imitate certain emotions or perform conversations to generate emotional reactions \cite{mansoorizadeh2010multimodal}. Asking someone to generate emotions while watching or listening something does not create the same authentic feeling. Real emotions, on the other hand, are hard to control and record in a laboratory setting. 

Collecting data is a challenging task because of the wide variability of possible inputs. Usually, the researcher uses a small set of data that is available for training and adds unlabeled data into it to make the classifier more robust. Adding of unlabeled data for training needs a lot more care. Probabilistic models are used to deal with this scenario \cite{Cohen:2015:SMI:2749359.2735589}. Recently, deep learning algorithms are used widely for multimodal input classification \cite{ngiam2011multimodal,baltruvsaitis2018challenges}.

\subsection{Evaluation Measures}
The main issue in MMIS design research is the evaluation. Various measures such as efficiency, quality, user satisfaction, and accuracy are available in the literature to evaluate an interface. Usually, computational measures, such as efficiency, are used because the interface can improve the interaction, task completion, and decrease the complexity of the system. Task completion time is one way to measure efficiency. Another measure is the number of actions the user has performed to decide or solve a problem \cite{remy2018evaluation}.

Another way to measure the effectiveness of an interface is to ask the subjective opinion of the user through questionnaires. The questionnaires can ask the user about user satisfaction, fatigue level, system compatibility, and easiness to use the system \cite{remy2018evaluation}. The biofeedback signals such as EEG, ECG, etc. can also be used to determine the system effectiveness by measure the user emotional level or cognitive state. 

 \section{Current Applications of MMIS}
In the previous sections, we have discussed various types of modalities, fusion models, data collection for training and testing, and evaluation of the interface. This section will discuss some of the recent applications of HCI in various scenarios including ambient spaces, mobile and wearable technology, virtual environments, and arts, etc.

Multimodal systems have been used in video indexing for detecting human behavior and expressions \cite{zeng2009survey,pantic2007human}. Virtual meeting rooms utilize multimodal systems to record and model user behavior in real-time \cite{reidsma2007virtual}. Behavioral analysis from multimodal input is utilized in surveillance and intelligence applications \cite{zhu2007multimodal,roth2003design}. Bradbury et al. \cite{bradbury2003hands} proposed an attentive kitchen concept named eyeCOOK, which combines eye-gaze and speech commands to help non-competent users cook a meal. Multimodal systems have been applied to various healthcare applications. Gestonurse is a multimodal robotic scrub nurse that help the chief surgeon by giving surgical equipment. The speech and gesture modalities are used as input \cite{jacob2012gestonurse}. Nie et al. \cite{nie2015beyond} present a health prediction system based on multimodal observation. Smart home concepts also utilize multimodal inputs to record various activities of the user \cite{fleury2010svm,brdiczka2009detecting}.

Smartphones are the most prominent example of multimodal interface systems. Today's smartphones have the capability to interact with speech, touch, gesture, gaze, and facial inputs \cite{shaheen2017smartphone}. Virtual and augmented reality apps open the ways to unlimited possibilities of applications, and with the multimodal inputs, these applications provide a very natural feeling of interaction \cite{bekele2018survey,billinghurst2015survey}. A real-time strategy game in which the agents can be instructed with speech and gesture commands \cite{link2016intelligent}. In another application, an augmented reality dialog interface has been present that enables the user to control the robot's verbal and non-verbal behavior \cite{pereira2017augmented} accurately. A method of designing intelligent interface for people with functional disabilities have been presented, which can combine visual, sound, and tactile multimodal inputs in a brain-computer interface to provide highly adaptable and personal services \cite{stirenko2017user}.

With the advancement in multimodal interaction and visual technology, the arts industry has produced some of the most exciting applications. Multimodal systems to play music \cite{wicaksono2017fabrickeyboard,ng20073d,overholt2009multimodal}, generate avatars \cite{lisetti2002maui,shaked2017avatars}, interacting with objects in museums \cite{liarokapis2017multimodal,capurro2015tangible} are some of the applications in the art industry. MMIS has been used to improve the lifestyle of people with disabilities. Speech-activated smart wheelchairs \cite{leaman2017comprehensive,munteanu2016designing}, brain-computer interfaces \cite{miralles2015brain,ferreira2013survey} and interfaces that use eye blinks or eyebrow movements \cite{lin2012wireless} are some examples that provide accessibility to disabled people.
MMIS is used in every aspect of computing that involves interaction with humans, objects, and environment. Although MMIS may not totally replace the traditional interaction systems, they have promising applications that can revolutionize the entertainment, games, art, and health-care industries. 

\section{Challenges in multimodal HCI}
Computing field has seen some significant developments in HCI in recent decades. The touch-based interfaces have become a vital part
in smartphones these days, and with multiple visual sensors on the device, researchers are continuously looking for new ways to interact with the device. Despite the advancements in technology, there are challenges and research problems that need addressing. Each modality itself is an active field of research such as gesture recognition, speech recognition, natural language understanding, activity recognition, haptics, user modeling, context understanding, etc. Deep learning algorithms have provided substantial support to the recognition algorithms, but much more research is needed in improving performance, personalization, integration, and adaptability.

In addition, we need to understand the user dependent issues that affect the performance of the multimodal interfaces. The study of the cognitive load of the user when interacting with an MMIS and the relationship of cognitive load with multiple modalities need further investigation \cite{chen2012multimodal}. 

\section{Conclusion}
In this paper, the research literature related to the multimodal interface systems has been presented. We presented an overview of multimodal interaction and history of some of the earliest MMIS. The MMIS provides the advantages to robustness, adaptability, improvement in task completion rate over a unimodal system. The evaluations show that multimodal interfaces improve the task completion rate by only 10\%  \cite{Dumas2009}, but in case of error handling and reliability, multimodal interfaces reduce errors by 36\% compared to unimodal interfaces.

Speech and gestures are the most widely used input modalities, along with touch and pen-based inputs. Most of the work in multimodal interaction is focused on input recognition such as gesture, speech, and facial expression recognition. A few studies focus on the output modality, channels of sensory output between human and computer, which is also a key element of human-computer interaction \cite{obrenovic2004modeling}. Microsoft Kinect and Leap motion sensors are more common for gesture recognition \cite{molchanov2016online}. Biofeedback devices are catching the researcher's interest because these devices provide an indirect representation of the user's emotional and cognitive states. The signals such as EEG, ECG, etc. can  determine the system effectiveness by measure the user emotional level or cognitive state.

Multimodal inputs are integrated at the signal, feature, and semantic levels. The recent applications have used semantic level integration because it is a late integration process which gives the advantage to update the modalities and vocabulary quickly. Data collection and testing are one of the most critical parts, and they require more attention because most of the time, data is collected in a controlled environment. The next important part is the evaluation of the interface, which can be qualitative and quantitative. In the qualitative evaluation, questionnaires are widely used. For qualitative analysis, the measures such as task completion rate, the average time taken to complete a task are used.



\section{Introduction}

\bibliographystyle{unsrt}  
\bibliography{references}  






\end{document}